\documentclass[lettersize,journal]{IEEEtran}
\usepackage{setspace}
\usepackage{amsmath}
\usepackage{amssymb}
\usepackage{amsfonts}
\usepackage{algorithm}
\usepackage{dsfont}
\usepackage{float}
\usepackage{cite}
\usepackage{graphicx}
\usepackage{epsfig}
\usepackage{subfigure}
\usepackage{psfrag}
\usepackage{xcolor}
\usepackage{url}
\usepackage{verbatim}
\usepackage{multirow}
\usepackage{subfigure}
\usepackage[colorlinks,linkcolor=black,anchorcolor=black,citecolor=black]{hyperref}

\def\BibTeX{{\rm B\kern-.05em{\sc i\kern-.025em b}\kern-.08em
    T\kern-.1667em\lower.7ex\hbox{E}\kern-.125emX}}

\begin{document}

\title{{Detecting Abrupt Change of Channel Covariance Matrix in IRS-Assisted Communication}
\thanks{
Manuscript received July 24, 2023; revised September 26, 2023, and October 18, 2023; accepted October 25, 2023.
The work of R. Liu, Y. Xu, D. He, and W. Zhang was supported in part by  National Natural Science Foundation of China Program (62371291, 62271316, 62101322), the Fundamental Research Funds for the Central Universities and Shanghai Key Laboratory of Digital Media Processing (STCSM 18DZ2270700).
The work of L. Liu and C. W. Chen was supported in part by the Research Grants Council, Hong Kong, China, under Grants 15203222 and 15213322.
The associate editor coordinating the review of this article and approving it for publication was Y. Zhu. \textit{(Corresponding
author: Liang Liu)}}
\thanks{R. Liu, Y. Xu, D. He and W. Zhang are with the Cooperative Medianet Innovation Center (CMIC), Shanghai Jiao Tong University, Shanghai, China (emails: {liurunnan, xuyin, hedazhi, zhangwenjun}@sjtu.edu.cn). R. Liu was also with the Hong Kong Polytechnic University, Hong Kong SAR, China.}
\thanks{L. Liu and C. W. Chen are with the Hong Kong Polytechnic University, Hong Kong SAR, China (e-mails: {liang-eie.liu, changwen.chen}@polyu.edu.hk).}
}

\author{Runnan Liu, Liang Liu, Yin Xu, Dazhi He, Wenjun Zhang, and Chang Wen Chen \vspace{-0.85cm} }

\maketitle

\begin{abstract}
The knowledge of channel covariance matrices is crucial to the design of intelligent reflecting surface (IRS) assisted communication.
However, channel covariance matrices may change suddenly in practice.
This letter focuses on the detection of the above change in IRS-assisted communication.
Specifically, we consider the uplink communication system consisting of a single-antenna user (UE), an IRS, and a multi-antenna base station (BS).
We first categorize two types of channel covariance matrix changes based on their impact on system design: Type I change, which denotes the change in the BS receive covariance matrix, and Type II change, which denotes the change in the IRS transmit/receive covariance matrix.
Secondly, a powerful method is proposed to detect whether a Type I change occurs, a Type II change occurs, or no change occurs.
The effectiveness of our proposed scheme is verified by numerical results.
\end{abstract}

\begin{IEEEkeywords}
Change detection, intelligent reflecting surface.
\end{IEEEkeywords}

\vspace{-10pt}

\section{Introduction}
Acquisition of the channel covariance matrices is of paramount importance to the performance in intelligent reflecting surface (IRS) assisted systems \cite{renzo2019smart,wu2021intelligent}. For example, the minimum mean-squared error (MMSE) estimator of a Rayleigh fading channel is a function of the channel covariance matrix. Moreover, to reduce the channel estimation overhead in the conventional IRS design \cite{wang2020channel,huang2019reconfigurable}, recently, plenty of works have utilized the channel covariance matrix to design the IRS reflecting coefficients for optimizing the long-term performance \cite{Jin19,zhao2020intelligent}. It is worth noting that despite slowly, the channel covariance matrices do change in practice due to the change in the scattering environment. In the literature of IRS-assisted systems, it remains an open problem in how to detect the change in the channel covariance matrices quickly and accurately such that we can re-estimate the new covariance matrices for updating the MMSE channel estimators, IRS reflecting coefficients, etc.

In this letter, we aim to tackle the above challenge in an IRS-assisted uplink communication system that consists of a single-antenna user, an IRS, and a multi-antenna base station (BS). Under this considered system, the BS receive covariance matrix, the IRS receive covariance matrix, and the IRS transmit covariance matrix can affect the system design in various ways. Specifically, our recent work \cite{wang2021massive} showed that the former covariance matrix merely affects the MMSE channel estimators, while the latter two have impact on both the MMSE channel estimators and the IRS reflecting coefficients. Therefore, we define the change in the BS receive covariance matrix as the Type I change, and that in the IRS receive covariance matrix and the IRS transmit covariance matrix as the Type II change. Then, based on the change detection theory \cite{basseville1993detection}, we design an efficient method that is able to 1) detect whether a change in channel covariance matrices occurs; and 2) if a change does occur, whether it is a Type I change or Type II change. 

It is worth noting that our recent work \cite{liu2022detecting} proposed an efficient algorithm to detect the change in the covariance matrix of a Rayleigh fading channel. However, this method cannot be applied in our considered IRS-assisted system. First, the cascaded channels in IRS-assisted communication are not Gaussian distributed. Second, \cite{liu2022detecting} only considers two hypotheses - change occurs and no change occurs, while our letter works for three hypotheses - Type I change occurs, Type II change occurs, and no change occurs.

\vspace{-10pt}

\section{System Model}\label{Sec:SysModel}
\begin{figure}
	\vspace{-0.1cm}
	\setlength{\abovecaptionskip}{-0.01 cm}
	\centering
	\includegraphics[width=9.7 cm]{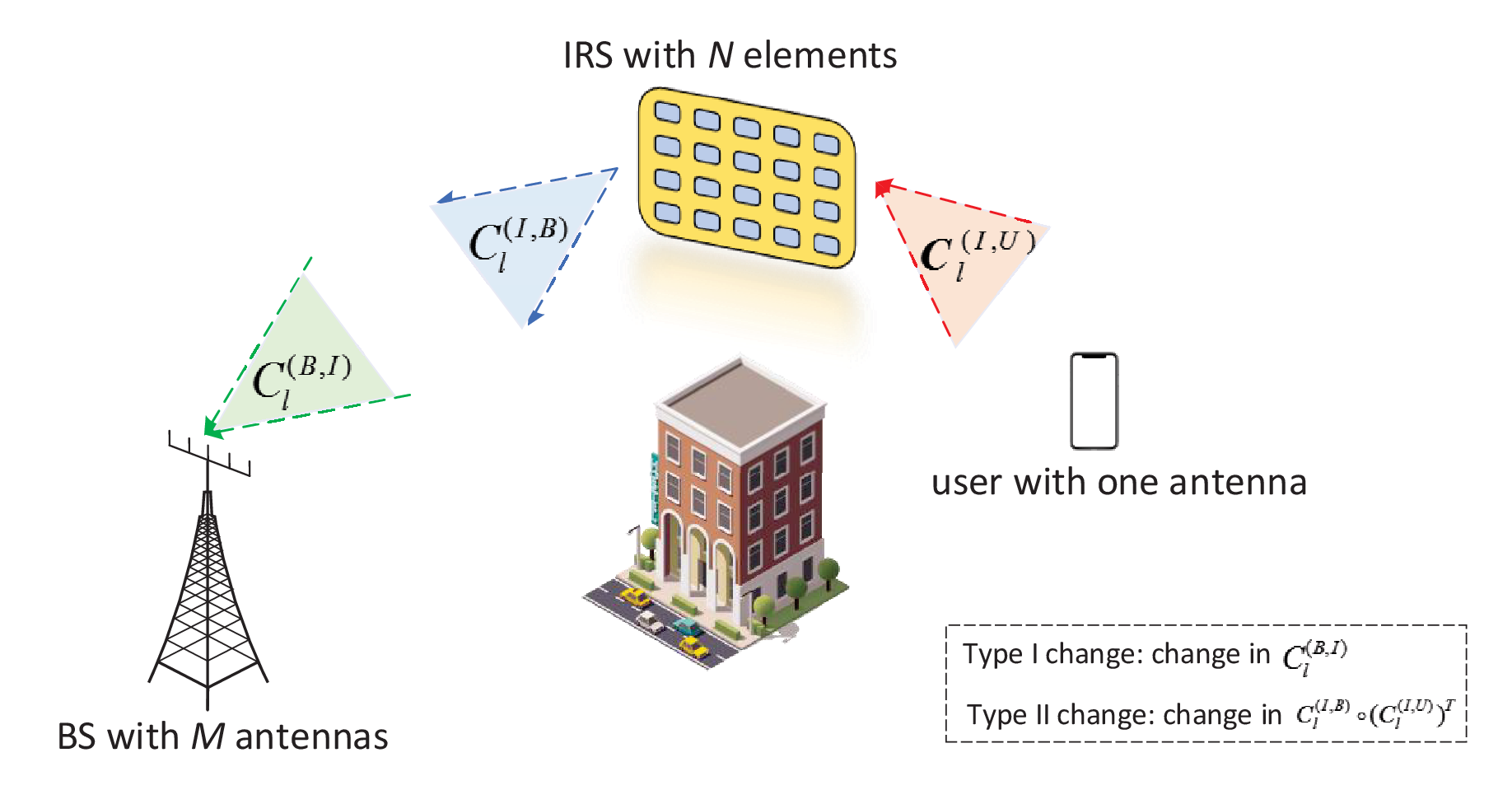}
	\caption{An IRS-assisted communication system.}
	\label{Fig:SysMod}
	\vspace{-0.4cm}
\end{figure}
Consider a narrow-band uplink communication system consisting of one BS equipped with $M$ antennas, one IRS equipped with $N$ IRS reflecting elements, and one single-antenna user, as shown in Fig. \ref{Fig:SysMod}.
In this letter, we assume that the values of $M$ and $N$ are very large, because the massive MIMO technique is mature in the current cellular system, and the number of IRS elements should be sufficiently large to compensate the path loss. Moreover, we consider a block fading channel model, where the channels stay constant in one coherence time interval, but may vary independently over different coherence time intervals.
Define $t_{l,n}$ and $\boldsymbol{r}_{l,n}$ as the channel from the user to the IRS reflecting element $n$ at coherence interval $l$ and that from IRS reflecting element $n$ to the BS at the coherence interval $l$, respectively.
Moreover, define $\boldsymbol{t}_l=[t_{l,1},\cdots,t_{l,N}]^T$ and $\boldsymbol{R}_l=[\boldsymbol{r}_{l,1},\cdots,\boldsymbol{r}_{l,N}]$, $\forall l$, for convenience.
Similar to \cite{wang2021massive}, we assume a Rayleigh fading channel model for $\boldsymbol{t}_l$ and $\boldsymbol{R}_l$.
Specifically, $\boldsymbol{t}_l \sim \mathcal{CN}(\boldsymbol{0}, \beta^{(\text{t})}_l\boldsymbol{C}^{(\text{I,U})}_l),~\forall l$, where $\beta^{(\text{t})}_l$ denotes the path loss of $\boldsymbol{t}_l$, and $\boldsymbol{C}^{(\text{I,U})}_l$ denotes the normalized IRS receive covariance matrix for the user with $\text{tr}(\boldsymbol{C}^{(\text{I,U})}_l)=N$.
Moreover, $\boldsymbol{R}_l$ has the same distribution as $(\boldsymbol{C}^{(\text{B,I})}_l)^{\frac{1}{2}}\tilde{\boldsymbol{R}_l}(\boldsymbol{C}^{(\text{I,B})}_l)^{\frac{1}{2}},~\forall l$, where $\text{vec}(\tilde{\boldsymbol{R}_l})\sim\mathcal{CN}(\boldsymbol{0},N\beta^{(\text{R})}_l\boldsymbol{I})$ with $\beta^{(\text{R})}_l$ denoting the path loss of $\boldsymbol{R}_l$, $\boldsymbol{C}^{(\text{B,I})}_l$ denotes the normalized BS receive covariance matrix for the IRS with $\text{tr}(\boldsymbol{C}^{(\text{B,I})}_l)=M$, and $\boldsymbol{C}^{(\text{I,B})}_l$ denotes the normalized IRS transmit covariance matrix for the BS with $\text{tr}(\boldsymbol{C}^{(\text{I,B})}_l)=N$.
Furthermore, we assume that the direct channel between the user and the BS is blocked by some obstacles and does not exist.
Define $x^{(\text{p})}_l$ with $|x^{(\text{p})}_l|=1$ and $\boldsymbol{y}^{(\text{p})}_l$ as the transmitted pilot signal and the received pilot signal at the channel estimation phase of coherence time interval $l$, while $x^{(\text{d})}_l\sim \mathcal{CN}(0,1)$ and $\boldsymbol{y}^{(\text{d})}_l$ as the transmitted data signal and the received data signal at the data transmission phase of coherence time interval $l$.
Then, the received signal of the BS at coherence time interval $l$ is given as
\begin{equation}
	\boldsymbol{y}^{(\Omega)}_l = \sqrt{w}\sum_{n=1}^{N}\phi_{l,n}t_{l,n}\boldsymbol{r}_{l,n}x^{(\Omega)}_l+\boldsymbol{z}_l=\sqrt{w}\boldsymbol{h}_lx^{(\Omega)}_l+\boldsymbol{z}_l,
	\label{eq:received signal}
\end{equation}
where $\Omega\in\{\text{p},\text{d}\}$ denotes the indicator for the pilot signal and the data signal, $w$ denotes the transmit power of the user, $\phi_{l,n}$ denotes the reflecting coefficient of IRS element $n$ at coherence time interval $l$, $\boldsymbol{z}_l\sim\mathcal{CN}(\boldsymbol{0},\sigma^2\boldsymbol{I})$ denotes the additive white Gaussian noise (AWGN) at the BS, and
\begin{equation}\label{eqn:effective channel}
    \boldsymbol{h}_l=\sum_{n=1}^{N}\phi_{l,n}t_{l,n}\boldsymbol{r}_{l,n}
\end{equation}
denotes the effective channel from the user to the BS through the IRS at coherence time interval $l$.
It can be shown that the mean of $\boldsymbol{h}_l$ is $\boldsymbol{0}$, and the covariance matrix of $\boldsymbol{h}_l$ is
\begin{equation}
    \boldsymbol{V}_l=\beta^{(\text{R})}_l\beta^{(\text{t})}_l\Big(\boldsymbol{\phi}_l^H\boldsymbol{C}^{(\text{I,B})}_l\circ({\boldsymbol{C}^{(\text{I,U})}_l})^T\boldsymbol{\phi}_l\Big)\boldsymbol{C}^{(\text{B,I})}_l,
	\label{eq:CovMTX}
\end{equation}
where $\circ$ denotes the Hadamard product and $\boldsymbol{\phi}_l=[\phi_{l,1},\cdots,\phi_{l,N}]^T$.

\section{Two Types of Change for Detection}
In \cite{wang2021massive}, an achievable rate expression for the system model considered in Section \ref{Sec:SysModel} was given under the so-called two-timescale protocol \cite{Jin19,zhao2020intelligent}. Specifically, the BS receive beamforming vector is optimized at each coherence time interval $l$ based on the estimation of $\boldsymbol{h}_l$, while the IRS beamforming vectors are optimized based on the channel covariance matrices. Note that $\boldsymbol{C}^{(\text{I,U})}_l$, $\boldsymbol{C}^{(\text{B,I})}_l$, and $\boldsymbol{C}^{(\text{I,B})}_l$ may stay constant over many $l$'s.
Under this protocol, it was rigorously shown in \cite{wang2021massive} that as $M$ and $N$ go to infinity with a fixed ratio, $\boldsymbol{h}_l$ tends to be Gaussian distributed, i.e., $\boldsymbol{h}_l\sim\mathcal{CN}(\boldsymbol{0},\boldsymbol{V}_l),~\forall l$.
As a result, in the channel estimation phase of coherence time interval $l$, i.e., $\Omega=\text{p}$ in (\ref{eq:received signal}), the MMSE estimator of $\boldsymbol{h}_l$ is
\begin{equation}
    \hat{\boldsymbol{h}}_l=\frac{(x^{(\text{p})}_l)^H}{\sqrt{w}}\boldsymbol{V}_l(\boldsymbol{V}_l+\frac{\sigma^2}{w}\boldsymbol{I})^{-1}\boldsymbol{y}^{(\text{p})}_l,~\forall l.
    \label{eq:MMSE}
\end{equation}
Then, in the data transmission phase of coherence time interval $l$, i.e., $\Omega=\text{d}$ in (\ref{eq:received signal}), the BS utilizes the estimated channel to design the maximal-ratio combining (MRC) beamforming vector for decoding the user messages as
\begin{equation}
    \tilde{x}^{(\text{d})}_l=(\hat{\boldsymbol{h}}_l)^H\boldsymbol{y}^{(\text{d})}_l.
    \label{eq:MRCBeamforming}
\end{equation}
\cite{wang2021massive} showed that the user achievable rate at coherence time interval $l$ is
\begin{equation}
	R_l= \log\left(1+\frac{w\beta^{(\text{R})}_l\beta^{(\text{t})}_l}{N\sigma^2}\Big(\boldsymbol{\phi}_l^H\boldsymbol{C}^{(\text{I,B})}_l\circ({\boldsymbol{C}^{(\text{I,U})}_l})^T\boldsymbol{\phi}_l\Big)\right).
	\label{eq:rate}
\end{equation}
A key finding is that the user rate merely depends on $\boldsymbol{C}^{(\text{I,B})}_l$ and $\boldsymbol{C}^{(\text{I,U})}_l$, but has nothing to do with $\boldsymbol{C}^{(\text{B,I})}_l$.

In the following, we define two types of change in the channel covariance matrix based on their impact on the design of the above two-timescale protocol.
{\bf Type I Change}: Change in $\boldsymbol{C}^{(\text{B,I})}_l$. If an abrupt change in $\boldsymbol{C}^{(\text{B,I})}_l$ occurs, we need to change the MMSE channel estimator in (\ref{eq:MMSE}) and MRC beamforming vector in (\ref{eq:MRCBeamforming}).
However, we do not need to re-design the IRS reflecting coefficients according to (\ref{eq:rate}).
Note that if a Type I change is detected, we only need to re-estimate the covariance matrix of the effective channel $\boldsymbol{h}_l$ based on some channel samples to design the new MMSE channel estimator (\ref{eq:MMSE}).
{\bf Type II Change}: Change in $\boldsymbol{C}^{(\text{I,B})}_l\circ(\boldsymbol{C}^{(\text{I,U})}_l)^T$. If an abrupt change in $\boldsymbol{C}^{(\text{I,B})}_l$ or $\boldsymbol{C}^{(\text{I,U})}_l$ occurs, we need to change the MMSE channel estimator in (\ref{eq:MMSE}), the MRC beamforming vector in (\ref{eq:MRCBeamforming}), and the IRS reflection coefficients according to (\ref{eq:rate}). Note that if a Type II change is detected, we have to re-estimate $\boldsymbol{C}^{(\text{I,B})}_l\circ({\boldsymbol{C}^{(\text{I,U})}_l)^T}$ to design the new IRS reflection coefficients based on (\ref{eq:rate}), which is challenging.

In the rest of this letter, we will propose an efficient change detector to detect the Type I and the Type II changes. Specifically, since the channel covariance matrix changes in a larger timescale compared to the coherence time interval, we define a covariance change detection (CCD) interval as the collection of $K$ coherence time intervals. Define $\boldsymbol{Y}_j^{(\text{p})}=[\boldsymbol{y}_{(j-1)K+1}^{(\text{p})},\cdots,\boldsymbol{y}_{jK}^{(\text{p})}]$ as the pilot signals received over $K$ coherence time intervals of CCD interval $j$, $\forall j$. At the end of each CCD interval $j$, our job is to detect whether a Type I change or Type II change occurs in this interval based on $\boldsymbol{Y}_j^{(\text{p})}$. Because change detection is merely performed at the end of each CCD interval, we assume that the IRS reflecting coefficients will stay constant throughout each CCD interval under the two-timescale protocol \cite{Jin19,zhao2020intelligent}, i.e., $\phi_{(j-1)K+1,n}=\cdots=\phi_{jK,n}=\bar{\phi}_{j,n}$, $\forall j$, where $\bar{\phi}_{j,n}$ denotes the constant reflecting coefficient of IRS element $n$ throughout CCD interval $j$. Define
\begin{equation}
	  \begin{aligned}
	     \theta_{(j-1)K+i}=\bar{\boldsymbol{\phi}}_j^H\boldsymbol{C}^{(\text{I,B})}_{(j-1)K+i}\circ&({\boldsymbol{C}^{(\text{I,U})}_{(j-1)K+i}})^T\bar{\boldsymbol{\phi}}_j,\\
        &i=1,\cdots,K, ~ \forall j,
    \end{aligned}
   \label{eq:CovPower}
\end{equation}

where $\bar{\boldsymbol{\phi}}_j=[\bar{\phi}_{j,1},\cdots,\bar{\phi}_{j,N}]^T$. Then, if $\boldsymbol{C}^{(\text{I,B})}_{(j-1)K+i}\circ({\boldsymbol{C}^{(\text{I,U})}_{(j-1)K+i}})^T$ changes at some coherence time interval $i$ of CCD interval $j$, then it is equivalent to the fact that $\theta_{(j-1)K+i}$ changes at some coherence time interval $i$ of CCD interval $j$. Because detecting the change in a scalar is much easier than detecting the change in a matrix, in the rest of this letter, we detect a Type II change via detecting the change in $\theta_{(j-1)K+i}$, instead of $\boldsymbol{C}^{(\text{I,B})}_{(j-1)K+i}\circ({\boldsymbol{C}^{(\text{I,U})}_{(j-1)K+i}})^T$.

In this letter, we assume that the duration of the CCD interval is well determined such that within each CCD interval $j$, at most one Type I change or Type II change occurs. Then, at each CCD interval $j$, our job is to utilize $\boldsymbol{Y}_j^{(\text{p})}$ to detect among the following three hypotheses
\begin{itemize}
    \item $H_0$: No Type I change and Type II change occur, i.e.,
    \begin{equation}
    \begin{aligned}
        & \boldsymbol{C}^{(\text{B,I})}_{(j-1)K+i}=\bar{\boldsymbol{C}}^{(\text{B,I})}_{j-1}, ~i=1,\cdots,K,\\
       \text{and} ~ & \theta_{(j-1)K+i}=\bar{\theta}_{j-1},~i=1,\cdots,K,
        \label{eq:H0}
    \end{aligned}
    \end{equation}
where $\bar{\boldsymbol{C}}^{(\text{B,I})}_{j-1}=\boldsymbol{C}^{(\text{B,I})}_{(j-1)K}$ and $\bar{\theta}_{j-1}=\theta_{(j-1)K}$.
\end{itemize}
\begin{itemize}
\item $H_1$: A Type I change occurs, i.e.,
\begin{equation}
    \begin{aligned}
        &\theta_{(j-1)K+i}=\bar{\theta}_{j-1},~i=1,\cdots,K, \\
       \text{and} ~ & \exists ~\nu^{(C)}_j<K ~ \text{such that} \\
            &\left\{
            \begin{aligned}
                & \boldsymbol{C}^{(\text{B,I})}_{(j-1)K+i}=\bar{\boldsymbol{C}}^{(\text{B,I})}_{j-1},\quad\quad\quad i=1,\cdots,\nu^{(C)}_j-1,\\
                &\boldsymbol{C}^{(\text{B,I})}_{(j-1)K+i}=\bar{\boldsymbol{C}}^{(\text{B,I})}_{j}\neq\bar{\boldsymbol{C}}^{(\text{B,I})}_{j-1},~i=\nu^{(C)}_j,\cdots,K,
            \end{aligned}
            \right.
    \end{aligned}
    \label{eq:H1}
\end{equation}
where $\bar{\boldsymbol{C}}^{(\text{B,I})}_{j}$ denotes the new covariance matrix after the change in CCD interval $j$.
\item $H_2$: A Type II change occurs, i.e.,
\begin{equation}
    \begin{aligned}
        &\boldsymbol{C}^{(\text{B,I})}_{(j-1)K+i}=\bar{\boldsymbol{C}}^{(\text{B,I})}_{j-1},~i=1,\cdots,K, \\
      \text{and} ~  & \exists ~\nu^{(\theta)}_j<K ~ \text{such that}  \\
            &\left\{
            \begin{aligned}
                &\theta_{(j-1)K+i}=\bar{\theta}_{j-1},\quad\quad~~i=1,\cdots, \nu^{(\theta)}_j-1,\\
                &\theta_{(j-1)K+i}=\bar{\theta}_{j}\neq\bar{\theta}_{j-1},~ i=\nu^{(\theta)}_j,\cdots, K,
            \end{aligned}
            \right.
    \end{aligned}
    \label{eq:H2}
\end{equation}
where $\bar{\theta}_{j}$ denotes the new scalar value after the change in CCD interval $j$.
\end{itemize}

\section{Change Detection Scheme}
In this section, we introduce how to detect among three hypotheses $H_0$, $H_1$, and $H_2$ at each CCD interval $j$. Specifically, after receiving $\boldsymbol{Y}_j^{(\text{p})}$ in CCD interval $j$, we first estimate the channels over all the $K$ coherence time intervals of this CCD interval, which are denoted as $\bar{\boldsymbol{H}}_j=[\bar{\boldsymbol{h}}_{(j-1)K+1},\cdots,\bar{\boldsymbol{h}}_{jK}]$, and then detect whether the channel distribution has changed based on the $K$ channel samples in $\bar{\boldsymbol{H}}_j$. First, we introduce how to estimate the channels.
Note that before implementing change detection at the end of each CCD interval, we do not know whether some channel covariance matrix has changed at some time instant or not.
As a result, we adopt the maximum likelihood (ML) technique to estimate the channel at each coherence time interval for change detection, because ML estimators are independent of the channel distribution.
Specifically, at the $i$-th coherence time interval of CCD interval $j$, the ML channel estimator is
\begin{equation}
    \begin{aligned}
        \bar{\boldsymbol{h}}_{(j-1)K+i}=\boldsymbol{h}_{(j-1)K+i}+\bar{\boldsymbol{z}}_{(j-1)K+i}, ~ i=1,\cdots,K,
    \end{aligned}
\label{eq:ML_ch}
\end{equation}where $\bar{\boldsymbol{z}}_{(j-1)K+i}\sim\mathcal{CN}\big(\boldsymbol{0},\frac{\sigma^2}{w}\boldsymbol{I}\big)$.
According to (\ref{eq:ML_ch}), we have
$\bar{\boldsymbol{h}}_{l}\sim\mathcal{CN}(\boldsymbol{0},\beta^{(\text{R})}_{l}\beta^{(\text{t})}_{l}\theta_{l}\boldsymbol{C}^{(\text{B,I})}_{l}+\frac{\sigma^2}{w}\boldsymbol{I})$, $\forall l$.
Therefore, given any $\boldsymbol{C}^{(\text{B,I})}_{(j-1)K+i}$ and $\theta_{(j-1)K+i}$, the conditional probability density function (PDF) of $\bar{\boldsymbol{h}}_{(j-1)K+i},~i\leq K,~\forall j,$ is given in (\ref{eq:PDF}) at the top of the next page.
\begin{figure*}[t]
    \normalsize
    \vspace{-0.5cm}
    \begin{equation}
        p(\bar{\boldsymbol{h}}_{(j-1)K+i}|\boldsymbol{C}^{(\text{B,I})}_{(j-1)K+i},\theta_{(j-1)K+i})=\frac{\exp\big(-\bar{\boldsymbol{h}}_{(j-1)K+i}^H(\beta^{(\text{R})}_{(j-1)K+i}\beta^{(\text{t})}_{(j-1)K+i}\theta_{(j-1)K+i}\boldsymbol{C}^{(\text{B,I})}_{(j-1)K+i}+\frac{\sigma^2}{w}\boldsymbol{I})^{-1}\bar{\boldsymbol{h}}_{(j-1)K+i}\big)}{\pi^M|\beta^{(\text{R})}_{(j-1)K+i}\beta^{(\text{t})}_{(j-1)K+i}\theta_{(j-1)K+i}\boldsymbol{C}^{(\text{B,I})}_{(j-1)K+i}+\frac{\sigma^2}{w}\boldsymbol{I}|}.
        \label{eq:PDF}
    \end{equation}
	\hrulefill
	\vspace{-0.2cm}
\end{figure*}

Next, we focus on detecting Type I/II change based on the estimated channels. Note that usually, change detection is between two hypotheses - a change exists and no change exists. However, here, we have to detect among three hypotheses, because a change may be a Type I change or a Type II change. The basic idea for change detection proposed in this letter is as follows. At each CCD interval $j$, in \emph{Step I}, we detect between Hypotheses $H_0$ and $H_1$ to check whether a Type I change occurs; while in \emph{Step II}, we detect between Hypotheses $H_0$ and $H_2$ to check whether a Type II change occurs. If Hypothesis $H_0$ is detected in both Steps I and II of CCD interval $j$, then we declare no change at this CCD interval. If Hypothesis $H_1$ ($H_0$) is detected in Step I but Hypothesis $H_0$ ($H_2$) is detected in Step II of CCD interval $j$, then we declare a Type I (Type II) change at this CCD interval. However, if Hypothesis $H_1$ is detected in Step I and Hypothesis $H_2$ is detected in Step II at CCD interval $j$, then we need to conduct \emph{Step III} - determining whether the change is a Type I change or a Type II change. In the rest of this section, we introduce how to implement Steps I, II, and III at each CCD interval, respectively.

\subsection{Step I: Detecting Type I Change}

Similar to \cite{liu2022detecting}, we adopt the log-likelihood ratio (LLR) based method to detect between Hypotheses $H_0$ and $H_1$ \cite{basseville1993detection}, where we assume that a Type II change does not occur in CCD interval $j$, i.e., $\theta_{(j-1)K+i}=\bar{\theta}_{j-1}$, $i=1,\cdots,K$. Under this method, if we guess that a Type I change occurs at coherence time interval $i$ in CCD interval $j$, then we can estimate the new covariance matrix after change, i.e., $\bar{\boldsymbol{C}}^{(\text{B,I})}_j$, based on the channels estimated from coherence time interval $i$ to the last coherence time interval of CCD interval $j$, i.e., $\bar{\boldsymbol{h}}_{(j-1)K+i},\cdots,\bar{\boldsymbol{h}}_{jK}$. Let us define $\hat{\boldsymbol{C}}_{i,j}^{(\text{B,I})}$ as this estimation, and we will show how to obtain it later. Therefore, if we guess that a Type I change occurs at coherence time interval $i$ in CCD interval $j$, it follows that $\boldsymbol{C}^{(\text{B,I})}_{(j-1)K+\nu}=\bar{\boldsymbol{C}}^{(\text{B,I})}_{j-1}$ when $\nu=1,\cdots,i-1$, and $\boldsymbol{C}^{(\text{B,I})}_{(j-1)K+\nu}=\hat{\boldsymbol{C}}_{i,j}^{(\text{B,I})}$ when $\nu=i,\cdots,K$. Then, the probability to observe  $\bar{\boldsymbol{h}}_{(j-1)K+1},\cdots,\bar{\boldsymbol{h}}_{jK}$ given the above event is
\begin{align}
p_{i,j}^{(H_1)} =&\prod_{\nu=1}^{i-1} p(\bar{\boldsymbol{h}}_{(j-1)K+\nu}|\bar{\boldsymbol{C}}^{(\text{B,I})}_{j-1},\bar{\theta}_{j-1})\nonumber \\
        \times &\prod_{\nu=i}^{K}p(\bar{\boldsymbol{h}}_{(j-1)K+\nu}|\hat{\boldsymbol{C}}^{(\text{B,I})}_{i,j},\bar{\theta}_{j-1}), ~  i=1,\cdots,K.
\end{align}Moreover, the probability to observe $\bar{\boldsymbol{h}}_{(j-1)K+1},\cdots,\bar{\boldsymbol{h}}_{jK}$ given the event that no change occurs in CCD interval $j$ is given as
\begin{align}
p_j^{(H_0)} =&\prod_{\nu=1}^{K} p(\bar{\boldsymbol{h}}_{(j-1)K+\nu}|\bar{\boldsymbol{C}}^{(\text{B,I})}_{j-1},\bar{\theta}_{j-1}).
\end{align}Therefore, the LLR between the event that a Type I change occurs at coherence time interval $i$ in CCD interval $j$ and the event that no Type I occurs in CCD interval $j$ is defined as
\begin{align}\label{eq:LLR1}
\boldsymbol{LLR}_{i,j}^{\text{C}}\!(\!\hat{\boldsymbol{C}}_{i,j}^{(\text{B,I})}\!)\!&\!=\!\log p_{i,j}^{(H_1)}-\log p_j^{(H_0)}\nonumber \\ &\!=\!\sum_{\nu=i}^{K}\log\!\left(\!\frac{p(\bar{\boldsymbol{h}}_{(j\!-\!1)K\!+\!\nu}|\hat{\boldsymbol{C}}^{(\text{B,I})}_{i,j},\bar{\theta}_{j\!-\!1})}{p(\bar{\boldsymbol{h}}_{(j\!-\!1)K\!+\!\nu}|\bar{\boldsymbol{C}}^{(\text{B,I})}_{j\!-\!1},\bar{\theta}_{j\!-\!1})}\!\right)\!.
\end{align}Define
\begin{equation}
\bar{i}_{j}^{(\text{C})} = \arg\max_{1\leq i\leq K} \boldsymbol{LLR}^{\text{C}}_{i,j}(\hat{\boldsymbol{C}}_{i,j}^{(\text{B,I})}),
\end{equation}as the coherence time interval with the maximum LLR value. Then, according to the standard change detection theory \cite{basseville1993detection}, the Type I change detector at CCD interval $j$ is
\begin{align}\label{eq:ChgDetStru}
\boldsymbol{LLR}^{\text{C}}_{\bar{i}_{j}^{(\text{C})},j}(\hat{\boldsymbol{C}}_{\bar{i}_{j}^{(\text{C})},j}^{(\text{B,I})}) \mathop{\lessgtr}^{H_0}_{H_1} \omega^{(\text{C})},~\forall j,
\end{align}where $\omega^{(\text{C})}$ is a threshold. In other words, a Type I change is declared for CCD interval $j$ if and only if the maximum value of $\boldsymbol{LLR}^{\text{C}}_{i,j}(\hat{\boldsymbol{C}}_{i,j}^{(\text{B,I})})$'s, $i=1,\cdots,K$, is larger than the threshold $\omega^{(\text{C})}$. Moreover, if a Type I change is declared, then $\bar{i}_{j}^{(\text{C})}$ is the estimated time when the change occurs.

The remaining issue is how to obtain $\hat{\boldsymbol{C}}_{i,j}^{(\text{B,I})}$ as an estimate of $\bar{\boldsymbol{C}}^{(\text{B,I})}_j$ based on $\bar{\boldsymbol{h}}_{(j-1)K+i}^{(\text{p})},\cdots,\bar{\boldsymbol{h}}_{jK}^{(\text{p})}$, $\forall i,j$. According to the change detector (\ref{eq:ChgDetStru}), the ML estimator $\hat{\boldsymbol{C}}_{i,j}^{(\text{B,I})}$ should be the optimal solution to the following problem
\begin{equation}
	\begin{aligned}
		\mathop{\textup{Maximize}}_{\hat{\boldsymbol{C}}^{(\text{B,I})}_{i,j}}&\quad \boldsymbol{LLR}^{\text{C}}_{i,j}(\hat{\boldsymbol{C}}_{i,j}^{(\text{B,I})})\\
		\textup{Subject to}&\quad \text{tr}(\hat{\boldsymbol{C}}^{(\text{B,I})}_{i,j})=M,\quad\xi_l\boldsymbol{I}\preceq\hat{\boldsymbol{C}}^{(\text{B,I})}_{i,j}\preceq\xi_u\boldsymbol{I},
	\end{aligned}
	\label{eq:ML_optimization}
\end{equation}where the first constraint is to ensure that the ML estimator is a normalized matrix, and $\xi_l>0$ and $\xi_u>\xi_l$ are the lower bound and upper bound for the eigenvalues of the estimator, respectively, to guarantee that the estimator is well-conditioned \cite{aubry2012maximum}.
To solve Problem (\ref{eq:ML_optimization}), we first define the sample covariance matrix of $\{\bar{\boldsymbol{h}}_{(j-1)K+i},\dots,\bar{\boldsymbol{h}}_{jK}\}$ as  $\boldsymbol{C}_{i,j}^{(\text{sam})}=\frac{1}{K-i+1}\sum_{\nu=i}^{K}\bar{\boldsymbol{h}}_{(j-1)K+\nu}\bar{\boldsymbol{h}}_{(j-1)K+\nu}^H$.
Then, define the eigenvalue decomposition (EVD) of $\boldsymbol{C}_{i,j}^{(\text{sam})}$ as $\boldsymbol{C}_{i,j}^{(\text{sam})}=\boldsymbol{\Phi}_{i,j}\text{diag}(\boldsymbol{\varphi}_{i,j})(\boldsymbol{\Phi}_{i,j})^H$,
where  $\boldsymbol{\varphi}_{i,j}=[\varphi_{1,i,j},\cdots,\varphi_{M,i,j}]^T$ and $\boldsymbol{\Phi}_{i,j}$ consist of the eigenvalues and eigenvectors of $\boldsymbol{C}_{i,j}^{(\text{sam})}$, respectively.
According to \cite{aubry2012maximum}, the optimal solution of Problem (\ref{eq:ML_optimization}) is
\begin{equation}
\hat{\boldsymbol{C}}^{(\text{B,I})}_{i,j}=\boldsymbol{\Phi}_{i,j}\text{diag}(\boldsymbol{\lambda}_{i,j})\boldsymbol{\Phi}_{i,j}^H-\frac{\sigma^2}{w\bar{\theta}_{j-1}}\boldsymbol{I},
	\label{eq:Solution_C_ML}
\end{equation}
where $\boldsymbol{\lambda}_{i,j}=[\lambda_{1,i,j},\dots,\lambda_{M,i,j}]^T$ is the solution to the following problem
\begin{equation}
	\mathop{\textup{Minimize}}_{\boldsymbol{\lambda}_{i,j}}\quad \sum\limits_{m=1}^{M}\log\lambda_{m,i,j}+\sum\limits_{m=1}^{M}\frac{\varphi_{m,i,j}}{\lambda_{m,i,j}}
	\label{Prob:eigen}
\end{equation}
$$
	\begin{aligned}
		&\textup{Subject to:}\\
		&\sum_{m=1}^{M}\lambda_{m,i,j}=M+\frac{\sigma^2M}{w\beta^{(\text{R})}_{(j-1)K+i}\beta^{(\text{t})}_{(j-1)K+i}\bar{\theta}_{j-1}},\\
		&\lambda_{m,i,j}\geq\xi_l+\frac{\sigma^2}{w\beta^{(\text{R})}_{(j-1)K+i}\beta^{(\text{t})}_{(j-1)K+i}\bar{\theta}_{j-1}},~\forall m,\\
		&\lambda_{m,i,j}\leq \xi_u+\frac{\sigma^2}{w\beta^{(\text{R})}_{(j-1)K+i}\beta^{(\text{t})}_{(j-1)K+i}\bar{\theta}_{j-1}}, ~\forall m.
	\end{aligned}
$$
Note that in the above problem, all constraints are convex. Moreover, the objective function is the sum of a concave function and a convex function. Therefore, we can adopt the concave-convex-procedure-based (CCCP-based) algorithm proposed in \cite{yuille2003concave} to efficiently obtain a sub-optimal solution to Problem (\ref{Prob:eigen}).
\vspace{-0.3cm}

\subsection{Step II: Detecting Type II Change}
Next, we introduce the LLR-based method to detect between Hypotheses $H_0$ and $H_2$, where we assume that a Type I change does not occur in each CCD interval $j$, i.e., $\boldsymbol{C}^{(\text{B,I})}_{(j-1)K+i}=\bar{\boldsymbol{C}}^{(\text{B,I})}_{j-1}$, $i=1,\cdots,K$. Define $\hat{\theta}_{i,j}$ as the estimation of $\bar{\theta}_j$ based on the channels estimated from coherence time interval $i$ to the last coherence time interval of CCD interval $j$, i.e., $\bar{\boldsymbol{h}}_{(j-1)K+i},\cdots,\bar{\boldsymbol{h}}_{jK}$.
We will show how to perform such an estimation later.
Similar to (\ref{eq:LLR1}), it can be shown that the LLR between the event that a Type II change occurs at coherence time interval $i$ in CCD interval $j$ and the event that no Type II change occurs in CCD interval $j$ is defined as
\begin{align}\label{eq:LLR2}
\boldsymbol{LLR}_{i,j}^{\theta}(\hat{\theta}_{i,j})\!=\!\sum_{\nu=i}^{K}\log\!\left(\!\frac{p(\bar{\boldsymbol{h}}_{(j-1)K+i}|\bar{\boldsymbol{C}}^{(\text{B,I})}_{j-1},\hat{\theta}_{i,j})}{p(\bar{\boldsymbol{h}}_{(j-1)K+i}|\bar{\boldsymbol{C}}^{(\text{B,I})}_{j-1},\bar{\theta}_{j-1})}\!\right)\!.
\end{align}Next, define the index of the coherence time interval with the maximum LLR value in CCD interval $j$ as
\begin{align}
\bar{i}_{j}^{(\theta)} = \arg\max_{1\leq i \leq K} \boldsymbol{LLR}_{i,j}^{\theta}(\hat{\theta}_{i,j}).
\end{align}Then, the Type II change detector is given as
\begin{equation}
\boldsymbol{LLR}_{\bar{i}_{j}^{(\theta)},j}^{\theta}(\hat{\theta}_{\bar{i}_{j}^{(\theta)},j}) \mathop{\lessgtr}^{H_0}_{H_2} \omega^{(\theta)},~\forall j,
	\label{ChgDetPw}
\end{equation}where $\omega^{(\theta)}$ is a threshold. In other words, a Type II change is declared for CCD interval $j$ if and only if the maximum value of $\boldsymbol{LLR}_{i,j}^{\theta}(\hat{\theta}_{i,j})$'s, $i=1,\cdots,K$, is larger than the threshold $\omega^{(\theta)}$. Moreover, if a Type II change is declared, then $\bar{i}_{j}^{(\theta)}$ is the estimated time when the change occurs.

At last, we show how to obtain $\hat{\theta}_{i,j}$ based on $\bar{\boldsymbol{h}}_{(j-1)K+i},\cdots,\bar{\boldsymbol{h}}_{jK}$. According to the change detector (\ref{ChgDetPw}), we can solve the following problem to obtain the ML estimator $\hat{\theta}_{i,j}$
\begin{equation}
	\begin{aligned}
		\mathop{\textup{Maximize}}_{\hat{\theta}_{i,j}}\quad \boldsymbol{LLR}_{i,j}^{\theta}(\hat{\theta}_{i,j}) ,\quad
		\textup{Subject to}\quad \hat{\theta}_{i,j}>0.
	\end{aligned}
	\label{Prob:ML_beta}
\end{equation}Similar to Problem (\ref{Prob:eigen}), the objective function in Problem (\ref{Prob:ML_beta}) can be expressed as the sum of a convex function and a concave function. Therefore, we can still use the CCCP-based algorithm proposed in \cite{yuille2003concave} to obtain a sub-optimal solution.

\vspace{-10pt}
\subsection{Step III: Resolving Confusion}
If the detector (\ref{eq:ChgDetStru}) declares a Type I change and the detector (\ref{ChgDetPw}) declares a Type II change at the same time, we should make a decision to determine whether the change is a Type I change or a Type II change. In other words, we need to detect between Hypotheses $H_1$ and $H_2$. This goal can be achieved via utilizing the standard LLR-based detection method. Specifically, at CCD interval $j$, the probabilities of observing $\bar{\boldsymbol{h}}_{(j-1)K+1},\cdots,\bar{\boldsymbol{h}}_{jK}$ under Hypotheses $H_1$ and $H_2$ are given as
\begin{equation}
	\vspace{-0.2cm}
	\begin{aligned}
		p_{j}^{(H_1)} =&\prod_{\nu=1}^{\bar{i}_{j}^{(\text{C})}-1} p(\bar{\boldsymbol{h}}_{(j-1)K+\nu}|\bar{\boldsymbol{C}}^{(\text{B,I})}_{j-1},\bar{\theta}_{j-1})\\
		&\times\prod_{\nu=\bar{i}_{j}^{(\text{C})}}^{K}p(\bar{\boldsymbol{h}}_{(j-1)K+\nu}|\hat{\boldsymbol{C}}^{(\text{B,I})}_{\bar{i}_{j}^{(\text{C})},j},\bar{\theta}_{j-1}),
	\end{aligned}
\end{equation}
\begin{equation}
	\begin{aligned}
		p_{j}^{(H_2)} =
		&\prod_{\nu=1}^{\bar{i}_{j}^{(\theta)}-1} p(\bar{\boldsymbol{h}}_{(j-1)K+\nu}|\bar{\boldsymbol{C}}^{(\text{B,I})}_{j-1},\bar{\theta}_{j-1})\\
		&\times\prod_{\nu=\bar{i}_{j}^{(\theta)}}^{K} p(\bar{\boldsymbol{h}}_{(j-1)K+\nu}|\bar{\boldsymbol{C}}^{(\text{B,I})}_{j-1},\hat{\theta}_{\bar{i}_{j}^{(\theta)},j}).
	\end{aligned}
\end{equation}
Then, at each CCD interval $j$, the LLR-based detector is given as
$
\log(p_{j}^{(H_1)})-\log(p_{j}^{(H_2)}) \mathop{\lessgtr}^{H_2}_{H_1} \omega^{(\text{I,II})},
$ where $\omega^{(\text{I,II})}$ is a threshold.

\vspace{-10pt}
\section{Numerical Results}
In this section, we present numerical results to verify the effectiveness of the proposed method to detect Type I and Type II changes in the IRS-assisted communications.
We assume that the BS has $M=32$ antennas and the IRS has $N=128$ reflecting elements.
For the covariance matrices $\boldsymbol{C}^{(\textup{B,I})}_l$, $\boldsymbol{C}^{(\textup{I,B})}_{l}$ and $\boldsymbol{C}^{(\textup{I,U})}_{l}$, we adopt the one-ring model as that used in our previous work \cite{liu2022detecting}.
Under this model, we assume that the change in $\boldsymbol{C}^{(\textup{B,I})}_{l}$, $\boldsymbol{C}^{(\textup{I,B})}_{l}$, and $\boldsymbol{C}^{(\textup{I,U})}_{l}$ is caused by the change in angle-of-departure (AOD) or the angle-of-arrival (AOA) of the corresponding paths, which is denoted by $\Delta\bar{\chi}$. Therefore, a larger $\Delta\bar{\chi}$ indicates a more significant Type I/II change. Moreover, we adopt the missed detection (a change occurs, but it is not detected) probability, denoted by $P_{MD}$, and false alarm (no change occurs, but a change is detected) probability, denoted by $P_{FA}$, as metrics to evaluate the performance of our method for detecting Type I and Type II changes.

\begin{figure}
	\vspace{-0.5cm}
	\centering
	\includegraphics[width=8 cm]{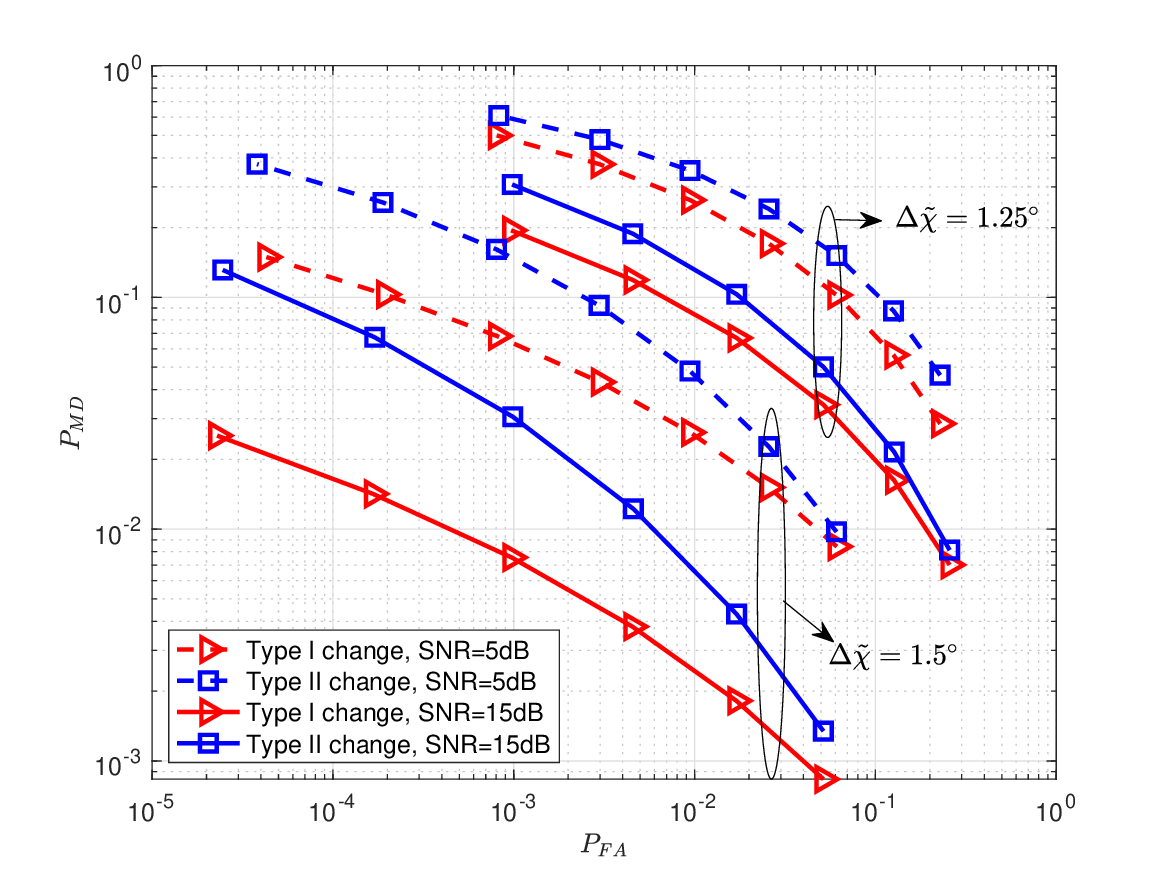}
	\caption{Trade-off between $P_{MD}$ and $P_{FA}$ for the proposed detection scheme.}
	\label{Fig_Result}
	\vspace{-0.4cm}
\end{figure}
Fig. \ref{Fig_Result} shows the trade-off between $P_{MD}$ and $P_{FA}$ for Type I change detection and Type II change detection when the signal-to-noise ratio (SNR) is 5 dB and 15 dB, $K=17$, and $\Delta\bar{\chi}=1.25^\circ, 1.5^\circ$.
First, it is observed that our proposed scheme can achieve high accuracy for detecting the Type I and Type II changes in IRS-assisted communication for both cases when SNR is 15 dB and 5 dB.
For example, for the case when SNR is 15 dB and the shift in AOA/AOD is $1.25^\circ$, with $P_{FA}=5\%$, $P_{MD}$ for Type I and that for Type II change detection are about $3.4\%$ and $4\%$. Second, when the shift in AOA/AOD is $1.5^\circ$ such that the change in the channel covariance matrix is more significant, the performance of our scheme is improved. For example, with $P_{FA}=0.45\%$, $P_{MD}$ for Type I and that for Type II change detection are about $0.38\%$ and $1.2\%$.

\vspace{-10pt}
\section{Conclusion}
In this letter, we defined the Type I change and the Type II change that can affect the design of the IRS-assisted communication in different manners. Then, based on the classic change detection theory, we proposed an efficient scheme to detect whether a Type I change occurs, a Type II change occurs, or no change occurs. Future work may consider how to re-estimate the channel covariance matrices after a  change is detected.

\bibliographystyle{IEEEtran}
\bibliography{Liu_WCL2023-1782}

\end{document}